\documentclass[aps,prb,10pt,twocolumn,floatfix,showpacs,superscriptaddress,longbibliography]{revtex4-1}

\usepackage{graphicx,amsmath,amsfonts,bm}
\usepackage[colorlinks=true, citecolor=blue, linkcolor=blue, urlcolor=blue]{hyperref}
\usepackage[all]{hypcap}
\usepackage{bbold}
\usepackage[normalem]{ulem}

\begin{document}

\title{Transport signatures of a Floquet topological transition at the helical edge}
\author{C. Fleckenstein}
\email{christoph.fleckenstein@physik.uni-wuerzburg.de}
\affiliation{Institute of Theoretical Physics and Astrophysics, University of W\"urzburg, 97074 W\"urzburg, Germany}
\author{N. Traverso Ziani}
\affiliation{Dipartimento di Fisica, Universit\`a di Genova, 16146 Genova, Italy}
\affiliation{CNR spin, 16146 Genova, Italy}
\author{L. Privitera}
\affiliation{Institute of Theoretical Physics and Astrophysics, University of W\"urzburg, 97074 W\"urzburg, Germany}
\author{M. Sassetti}
\affiliation{Dipartimento di Fisica, Universit\`a di Genova, 16146 Genova, Italy}
\affiliation{CNR spin, 16146 Genova, Italy}
\author{B. Trauzettel}
\affiliation{Institute of Theoretical Physics and Astrophysics, University of W\"urzburg, 97074 W\"urzburg, Germany}
\affiliation{W\"urzburg-Dresden Cluster of Excellence ct.qmat, Germany}
\begin{abstract}
Recently, the experimental realization of a quantum point contact (QPC) between helical edges has been accomplished. We predict that such a QPC, in the presence of a time periodic applied electric field, is characterized by a topological quantum phase transition in the Floquet spectrum. Crucially, the fact that the QPC is naturally attached to helical leads allows for the unambiguous detection of the topological Floquet quantum phase transition by means of multi-terminal transport experiments. Moreover, such experiments enable the characterization of the topological bound states associated with the transition.
\end{abstract}

\maketitle

\textit{Introduction.--} 
Since their theoretical prediction\cite{kane1,bernevig2006quantum} and experimental realization\cite{konig2007quantum}, two-dimensional topological insulators have triggered intense research activities in view of their potential applications in spintronics\cite{spintr1,spintr2,supspintr1,supspintr2} and topological quantum computation\cite{tqc1,tqc2}. Such applications require the ability to manipulate the edge states on demand. Several possibilities have been proposed at the level of a single edge. Proximity induced superconductivity, crucial for Majorana fermions, has already been experimentally demonstrated\cite{sc1}. Interaction induced gaps at the level of single edges, that could give rise to interesting fractional correlated phases\cite{fract1,fract2}, and in combination with superconductivity, to parafermions\cite{pf1,pf2}, are demanding because of the rather efficient screening in topological materials.

When two helical edges can interact with each other, the system becomes much richer\cite{teokane,nagaosa}. Recently, this route was substantiated by the experimental realization of the first quantum point contact (QPC) between two helical edges\cite{strunz}. This achievement could lead to the implementation of predicted phenomena relevant for spintronics applications\cite{qpc1,qpc2,qpc5,qpc4,Fleckenstein_PRB18}. Moreover, in the presence of both QPCs and superconductivity, Majorana fermions\cite{bernem} and parafermions\cite{para} can emerge.

\begin{figure}
\centering
\includegraphics[scale=0.225]{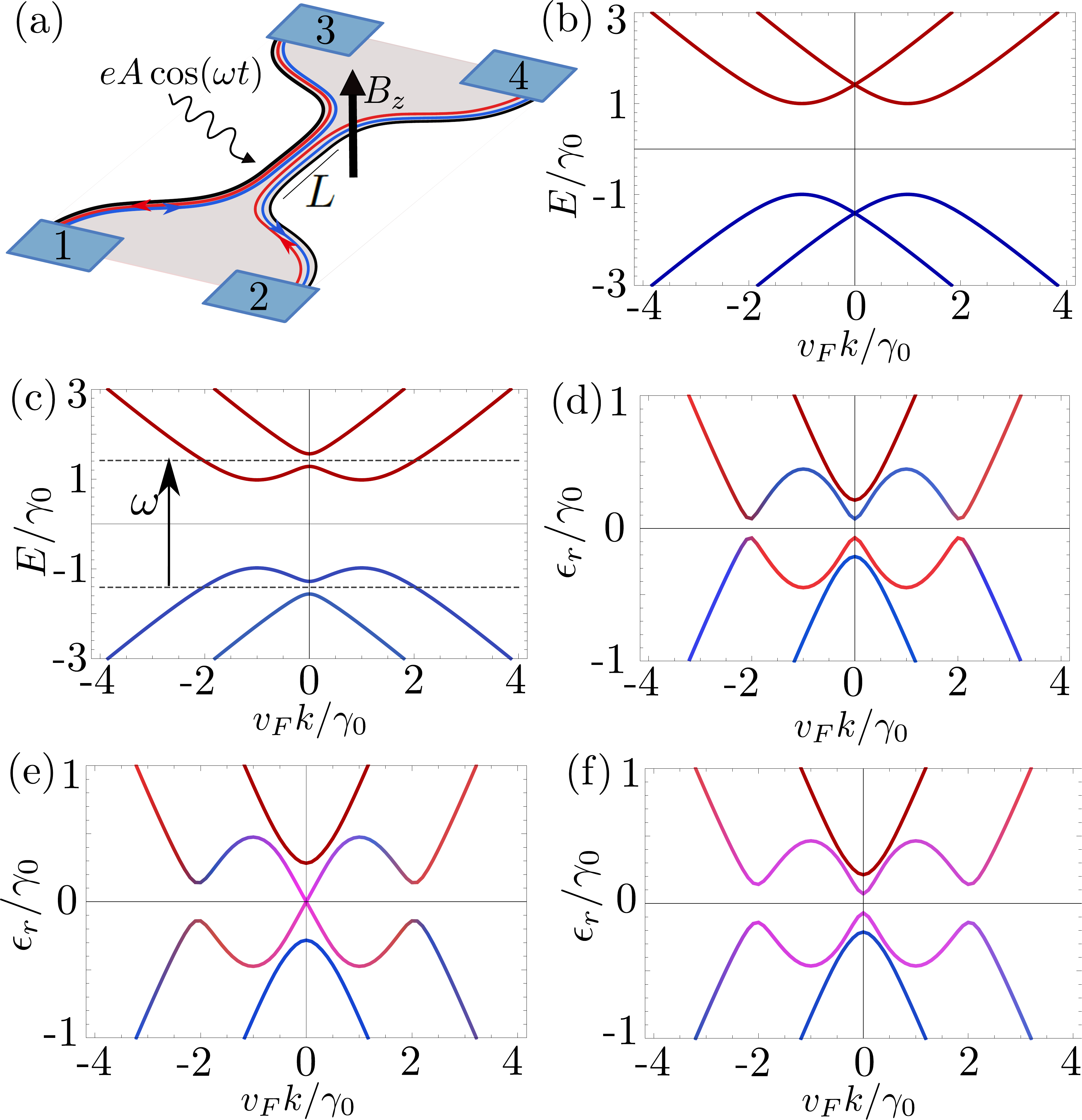}
\caption{(a) Schematic of the system: QPC based on a QSHI under the influence of electromagnetic radiation and Zeeman field. (b) QPC eigenvalue spectrum including all TR invariant static perturbations. (c) QPC eigenvalue spectrum with additional Zeeman field. The coupling induced by external electromagnetic radiation with resonant frequency $\omega=2\sqrt{\gamma_c^2+\gamma_0^2}$ ($\gamma_0=\gamma_c$) is also shown by means of the arrow. (d)-(f) Quasi-energy spectrum $\epsilon_r=\epsilon-\omega/2$ of the effective quasi-energy operator based on the extended Floquet-Hilbert space with (d) $eA/B_z=0.5$, (e) $eA/B_z=1$, (f) $eA/B_z=2$. The topological phase is characterized by  $0<eA/B_z <1$. The colors in the spectra (b-f) functions as a guide to the eye to visualize band inversion.}
\label{Fig:1}
\end{figure}

Time periodic (Floquet) external perturbations have been shown to be a powerful tool to engineer on demand particular properties~\cite{Yao_PRL07,Liu_Nat12,Sie_NMat14,floquetloss,kl1,Cavalleri_CP18,Oka_ARCMP19,Rudner_Nphys19,Mciver_arxiv18}. Likewise, the possibility of inducing topological phase transitions by means of driving has been analyzed in different contexts~\cite{oka2009photovoltaic,kitagawa2010topological,lindner2011floquet,kitagawa2011transport,Gu_PRL11,Liu_PRL13,Ezawa_PRL13,kundu,Rui_EPL14,141,Sentef_Naturecomm15,151,161,Privitera_PRB16,Yan_PRL16,Huebener_Ncomm17,Ezawa_PRB17,181,Rodriguez-Vega_PRB19,kl2}. Despite success in ultracold atomic gases~\cite{Jotzu_Nature14,Asteria_Nphys19,Tarnowski_Ncomm19}, a clear experimental signature of such phase transitions in solid-state systems is still lacking. Moreover, the most striking feature of a Floquet topological phase, that is, the appearance of topologically protected Floquet bound states, still needs to be experimentally verified.

In this work, we exploit the possibilities offered by the nanostucturing of the helical edges in combination with Floquet driving to propose a setup that can allow for a solution to both issues. We show that a periodically driven QPC between helical edges is characterized by a Floquet topological phase transition as the amplitude of the driving is varied. We demonstrate that the Floquet topological phase transition, associated with the presence of topological boundary states, can be unambiguously detected by photon-assisted transport. Indeed, both for weak and strong coupling to the leads, the conductance shows clear signatures of the phase transition. The transport properties are intimately connected with the spin properties of the bound state that emerges in the topological phase.

\textit{Model.--} The system under investigation is a QPC based on a quantum spin Hall insulator (QSHI) (Fig. \ref{Fig:1} (a)). The low energy physics, when all energy scales are smaller than the bulk gap, can be described in terms of four Fermi fields, collected in the spinor $\Psi(x)=[\hat{\psi}_{R,1}(x),\hat{\psi}_{L,1}(x),\hat{\psi}_{L,2}(x),\hat{\psi}_{R,2}(x)]^T$, where $R,L$ denote right- and left-movers and $1,2$ refer to the corresponding edge. We choose the spin $z$ projection such that $R,1$ corresponds to spin up, the other components follow accordingly. The kinetic energy is given by ($\hbar=1$)
\begin{equation}
H_p=\int\mathrm{d}x \Psi^{\dagger}(x)[-iv_F\partial_x \tau_z\sigma_z]\Psi(x),
\end{equation}
where $\tau_i$ and $\sigma_i$ with $i\in\{x,y,z\}$ are Pauli matrices acting on edge- and spin-space, respectively. This term characterizes both the QPC of length $L$, and the helical liquid leads. Assuming time-reversal symmetry (at zero magnetic field) we include two different scattering mechanisms \cite{bernem,Fleckenstein_PRB18}
\begin{eqnarray}
\label{Eq:Ht0}
H_{\gamma_0}&=&\gamma_0\int_0^L\mathrm{d}x\Psi^{\dagger}(x)\tau_x\sigma_0\Psi(x),\\
\label{Eq:Htc}
H_{\gamma_c}&=&\gamma_c\int_0^L\mathrm{d}x\Psi^{\dagger}(x)\tau_y\sigma_y\Psi(x),
\end{eqnarray} 
in the QPC region. While Eq. (\ref{Eq:Ht0}) describes the hybridization of states associated with different edges and does not need further symmetry breaking with respect to $H_p$, Eq. (\ref{Eq:Htc}) requires broken axial spin symmetry. The single-particle eigenvalue spectrum associated with $H_{\mathrm{QPC}}=H_p+H_{\gamma_0}+H_{\gamma_c}$ is shown in Fig. \ref{Fig:1} (b). We additionally apply a Zeeman field
\begin{eqnarray}
\label{Eq:HB}
H_B=B_z\int_{0}^L\mathrm{d}x \Psi^{\dagger}(x)\tau_0\sigma_z\Psi(x).
\end{eqnarray}
Then, a so-called helical gap around $k=0$ emerges (Fig. \ref{Fig:1} (c)).

The presence of a natural charge conjugation symmetry in our system, suggests the possibility to engineer topological properties related to this symmetry with a periodically driven external electro-magnetic field. The Hamiltonian associated to the driving is
\begin{eqnarray}
H_A(t)=\int_{0}^L\mathrm{d}x\Psi^{\dagger}(x)(-eA\cos(\omega t)\tau_z\sigma_z)\Psi(x).
\end{eqnarray}
When the frequency is chosen resonantly (i.e. $\omega=2\sqrt{\gamma_c^2+\gamma_0^2}$), the effective quasi-energy operator, derived below (Eq. (\ref{Eq:Heff})), undergoes a phase transition for $eA=B_z$ (Fig. \ref{Fig:1} (d)-(f)). To understand this dynamical transition in more detail, we apply the Shirley-Floquet theory to lowest order in the electric field keeping only one photon processes (see the Supplementary Material (SM)). Within this approximation, the quasi-energy operator $Q=H(t)-i\partial_t$ defining the eigenvalue problem for the time dependent states can be projected to the zero and one photon processes and assumes the form
\begin{equation}
\label{Eq:Heff}
H_{\mathrm{eff}}=\!\sum_k\!\textbf{d}_k^{\dagger}\!\begin{pmatrix}
\!E(-k)\! & \!B_{0}(k)\! & \!\Delta_0(-k) \!& \!0 \\
\!B_{0}(k)\! &\! E(k) \!& \!0 \!& \!\Delta_0(k) \\
\!\Delta_0(-k)\! &\! 0\! &\! -E(-k)\!+\!\omega\! &\! B_{0}(k)\\
\!0\! &\! \Delta_0(k) \!&\! B_{0}(k) \!&\! -E(k)\!+\!\omega \\
\end{pmatrix}\!\textbf{d}_k
\end{equation}
The new creation operators $\textbf{d}_k$ are associated with the eigenvalues $\pm E(\pm k)=\pm \sqrt{\gamma_0^2+(\gamma_c\pm kv_F)^2}$ (Fig. \ref{Fig:1} (b)), while the explicit expression of $B_0(k)\propto B_z$ is too lengthy to be reported here. \textcolor{black}{$\Delta_0(k)\propto~eA$ is explicitly given in the SM}.
Remarkably, around $eA=B_z$, we can find a gap-closing and reopening, indicating a topological phase transition in the Floquet spectrum (Fig. \ref{Fig:1} (d)-(f))\cite{note}. 

{The above phase transition happens at the dynamically generated charge-conjugation symmetric points at quasi-energy $\epsilon=\omega/2$. If the driving is not on resonance, the transition is slightly shifted.} 

\textit{Transport.--} For a structure shown in Fig. \ref{Fig:1} (a), the phase transition derived above can be detected by means of photon-assisted transport.  
To explain this detection scheme, we consider helical liquid leads coupled to the QPC. For simplicity, we assume the potentials in the constriction to be step functions $\gamma_0(x)=\gamma_c(x)=\gamma\theta(x)\theta(L-x)$ with the Heavyside function $\theta$\cite{note2}.
To access the conductance of such a heterostructure, we have to solve the corresponding Floquet scattering problem\cite{Moskalets_book}. In particular we solve the eigenvalue problem $Q(x)\vert \nu(x)\rangle=\epsilon\vert\nu(x)\rangle$ by computing the transfer matrix. We can rewrite the eigenvalue problem as
\begin{eqnarray}
\label{Eq:eig_prob}
P\partial_x\vert \nu(x)\rangle =\frac{i}{v_F}(\epsilon-Q_1(x))\vert\nu(x)\rangle
\end{eqnarray}
with $P=\mathbb{1}_{\infty}\tau_z\sigma_z$ and $Q_1=Q+iv_FP\partial_x$. We find the (infinite dimensional) transfer matrix of the QPC as
\begin{eqnarray}
M(L,0)=\exp\left[\int_0^L\mathrm{d}x\frac{i}{v_F}
P^{-1}(\epsilon-Q_1(x))\right].
\end{eqnarray}
An incoming mode can be scattered in any sector with photon number $m$ and edge index $j=1,2$. However, transitions including high numbers of photons are considerably less probable and the reflection-/transmission-amplitudes $r_{j}^m$, $t_{j}^m$ vanish with increasing photon number $m$. This allows us to compute the conductance with only a finite number $\overline{m}$ of harmonics by extracting the matrix $M_{\overline{m}}(L,0)=\exp\left[i/v_F\int_0^L\mathrm{d}x~P_{\overline{m}}^{-1}(\epsilon-Q_{1,\overline{m}})\right]$ out of the infinite dimensional matrix $M$. Indeed the transfer matrix $M_{\overline{m}}(L,0)$ defines a $4(2{\overline{m}+1})$ dimensional scattering problem that can be solved with appropriate incoming and outgoing states $\psi_{\mathrm{in,\overline{m}}}(x)$, $\psi_{\mathrm{out,\overline{m}}}(x)$, given in the SM.
\begin{figure}
\includegraphics[scale=0.33]{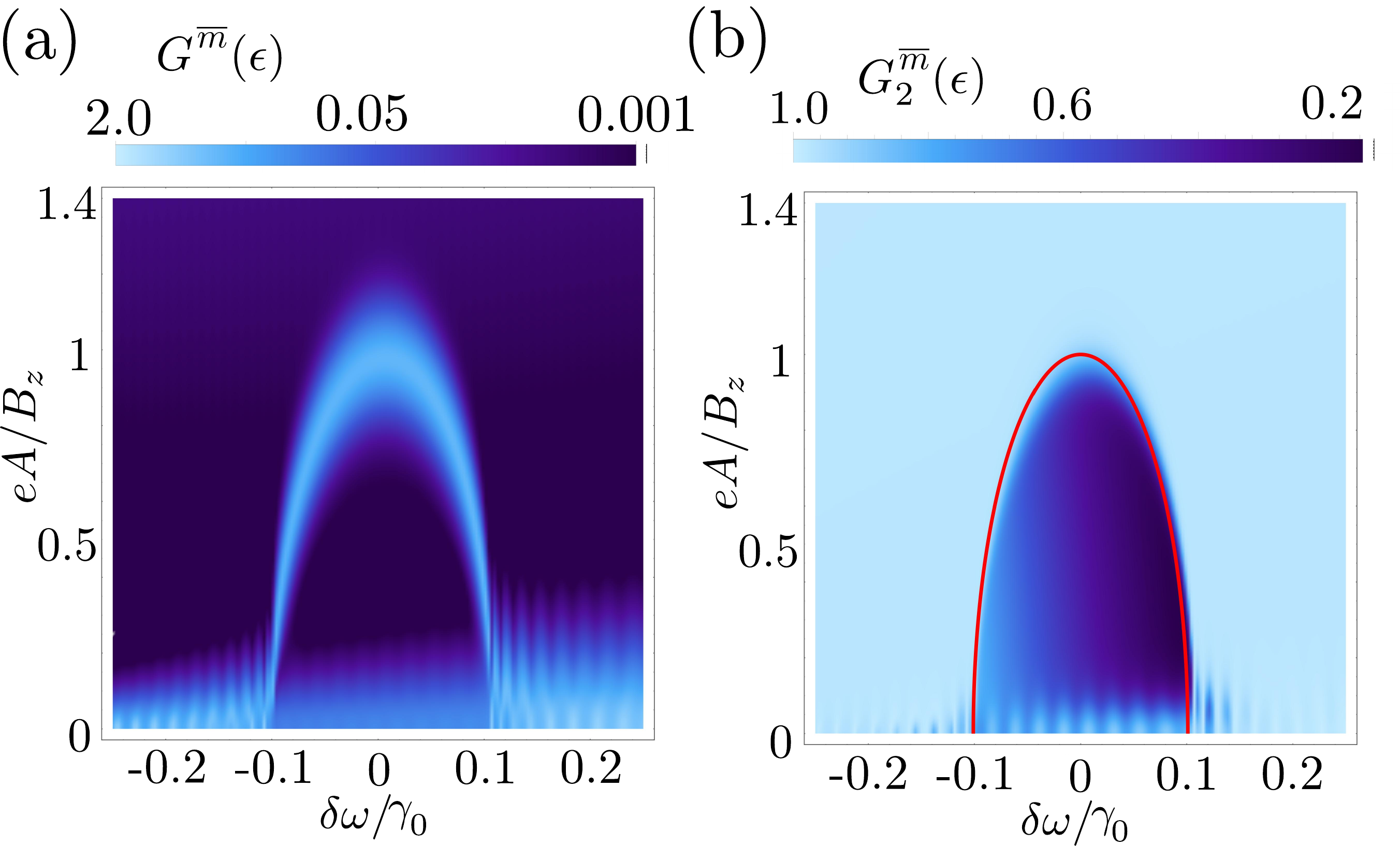}
\caption{Two terminal conductance evaluated from Eq. (\ref{Eq:cond_1}) (a) and four terminal conductance from Eq. (\ref{Eq:Conductance}) (b) in units of $e^2/h$ for quasi-energy $\epsilon=\sqrt{\gamma_0^2+\gamma_c^2}-\delta\omega$ and driving frequency $\omega=2(\sqrt{\gamma_0^2+\gamma_c^2}-\delta\omega)$. Additional parameters are $L=27~ v_F/\gamma_0$, $\overline{m}=10$, $v_F=1$ and $\gamma_0=\gamma_c$. The red line corresponds to the gap-closing-reopening of the effective quasi-energy operator (Eq. (\ref{Eq:GapClosing})) indicating the topological phase transition.}
\label{Fig:transport}
\end{figure}

Since the periodic drive is restricted to the QPC region, we are able to associate a Fermi distribution to each lead. The direct current in lead $\alpha$ is then given by \cite{tg,Moskalets_book,Moskalets_PRB02, Bardarson2017}
\begin{eqnarray}
\label{Eq:current}
I_{\alpha}&=&\frac{e}{2\pi}\int_{0}^{\infty}\mathrm{d}E \sum_{m=-\infty}^{\infty}\sum_{\beta}\bigg[\vert S_{\alpha,\beta}(E+m\omega,E)\vert^2 f_{\beta}(E)\nonumber\\
&-&\vert S_{\beta,\alpha}(E+m\omega,E)\vert^2 f_{\alpha}(E)\bigg]
\end{eqnarray}
with the Fermi distribution functions $f_{\alpha}(E)$ in lead $\alpha$ and the scattering matrix elements $S_{\alpha,\beta}(E+m\omega,E)$ capturing all scattering amplitudes for a transition from lead $\alpha$ to lead $\beta$ with photon number $m$. Assuming a small voltage difference $eV_\alpha$ between lead $\alpha$ and all other leads, and using the spatial-inversion symmetry of our scatterer (i.e. $\vert S_{\alpha,\beta}(E+m\omega,E)\vert^2=\vert S_{\beta,\alpha}(E+m\omega,E)\vert^2$), we obtain the linear conductance in lead $\alpha$ at zero temperature to $\overline{m}$-th order in the photon number
\begin{eqnarray}
\label{Eq:Conductance}
G^{\overline{m}}_{\alpha}(\epsilon)=\frac{e^2}{h}\big[1-\sum_{m=-\overline{m}}^{\overline{m}}\vert r_{\alpha}^{m}(\epsilon)\vert^2\big]
\end{eqnarray}
with the reflection amplitudes $r_{\alpha}^{m}(\epsilon)=r_{\alpha}(\epsilon+m\omega,\epsilon)$. \textcolor{black}{In the conductances we restore $h$}. From Eq. (\ref{Eq:Conductance}) we can directly derive the two terminal conductance\cite{note3}
\begin{eqnarray}
\label{Eq:cond_1}
G^{\overline{m}}(\epsilon)=\sum_{\alpha=1,2}G_{\alpha}^{\overline{m}}(\epsilon).
\end{eqnarray}
The results for the four-terminal ($G_{\alpha}^{\overline{m}}(\epsilon)$) as well as the two terminal conductance ($G^{\overline{m}}(\epsilon)$) are shown in Fig. \ref{Fig:transport} (a)-(b). Both are evaluated for $\epsilon_r=0$ with $\epsilon_r=\epsilon-\omega/2$ and $\omega=2(\sqrt{\gamma_0^2+\gamma_c^2}-\delta\omega)$.

Finite driving amplitude $eA$ blocks the two terminal transport through the QPC. This results in to zero conductance for sufficiently large QPC potentials (Fig. \ref{Fig:transport} (a)). In that case, finite conductance is only restored close to the phase transition (see the light-blue curve). In case of a four terminal measurement, the two phases are more prominently separated: In the trivial phase, the conductance is fixed to $e^2/h$ per channel, which implies the absence of backscattering in the channel of the incoming state. In this regime, time reversal symmetric processes dominate the physics and backscattering in the same channel is strongly suppressed. When the system undergoes the Floquet topological transition by crossing the red line of Fig. \ref{Fig:transport}, perfect conductance quantization is lost since the scattering is dominated by the TR breaking Zeeman field. Then, the conductance significantly differs from $e^2/h$. This unusual reflection is a direct consequence of a topological boundary state forming at the ends of the QPC at quasi energy $\epsilon =\omega/2$. 
\textcolor{black}{The oscillating pattern appearing in both plots for small $A$ can be explained by Fabry-P\'erot resonances emerging from interface scattering of the low energy wavefunctions.}

\begin{figure}[h]
	\centering
	\includegraphics[scale=0.3]{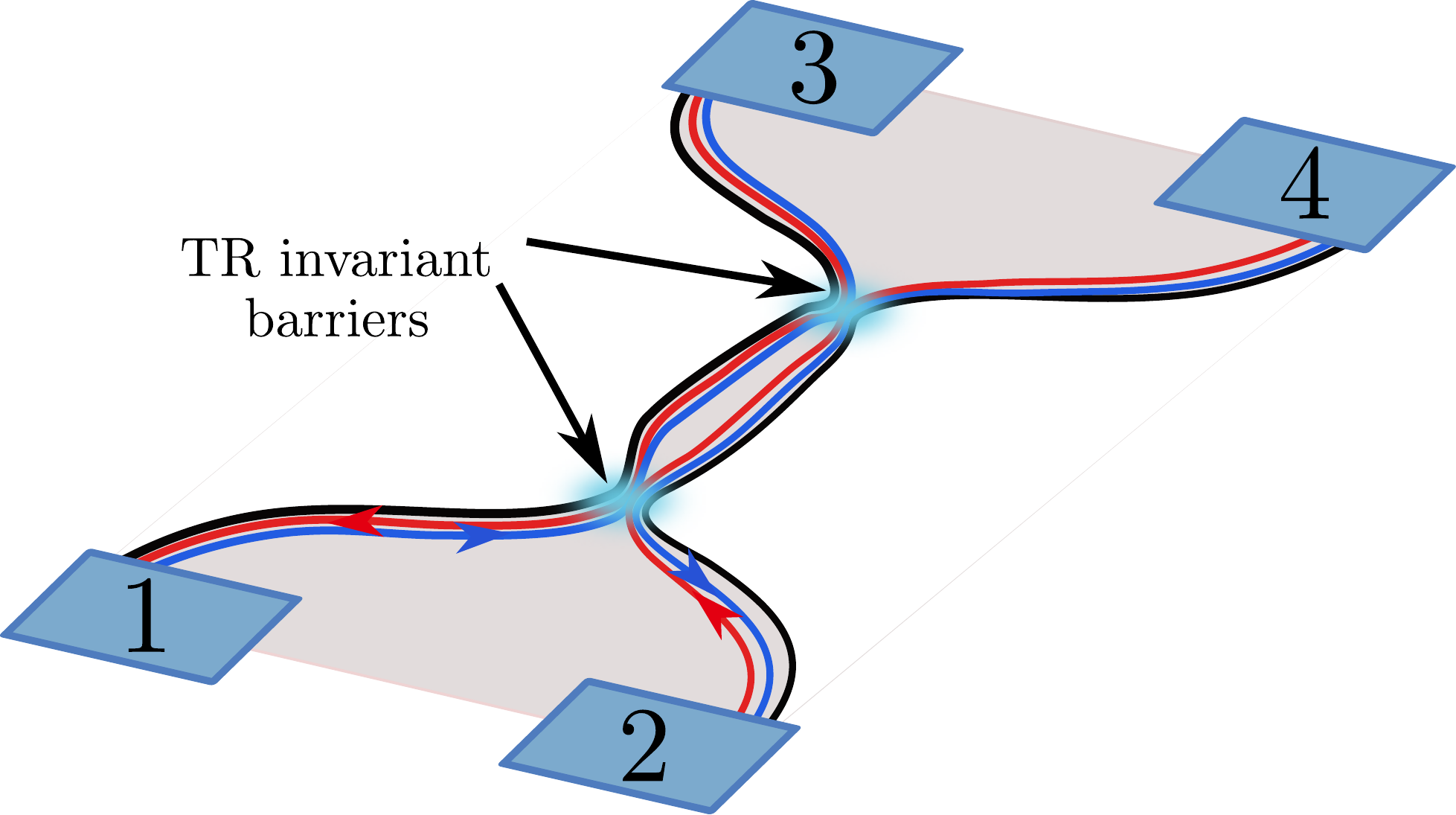}
	\label{Fig:schematic_supp}
	\caption{Schematic of a QPC with barriers.}
\end{figure}

\begin{figure}[h]
	\centering
	\includegraphics[scale=0.3]{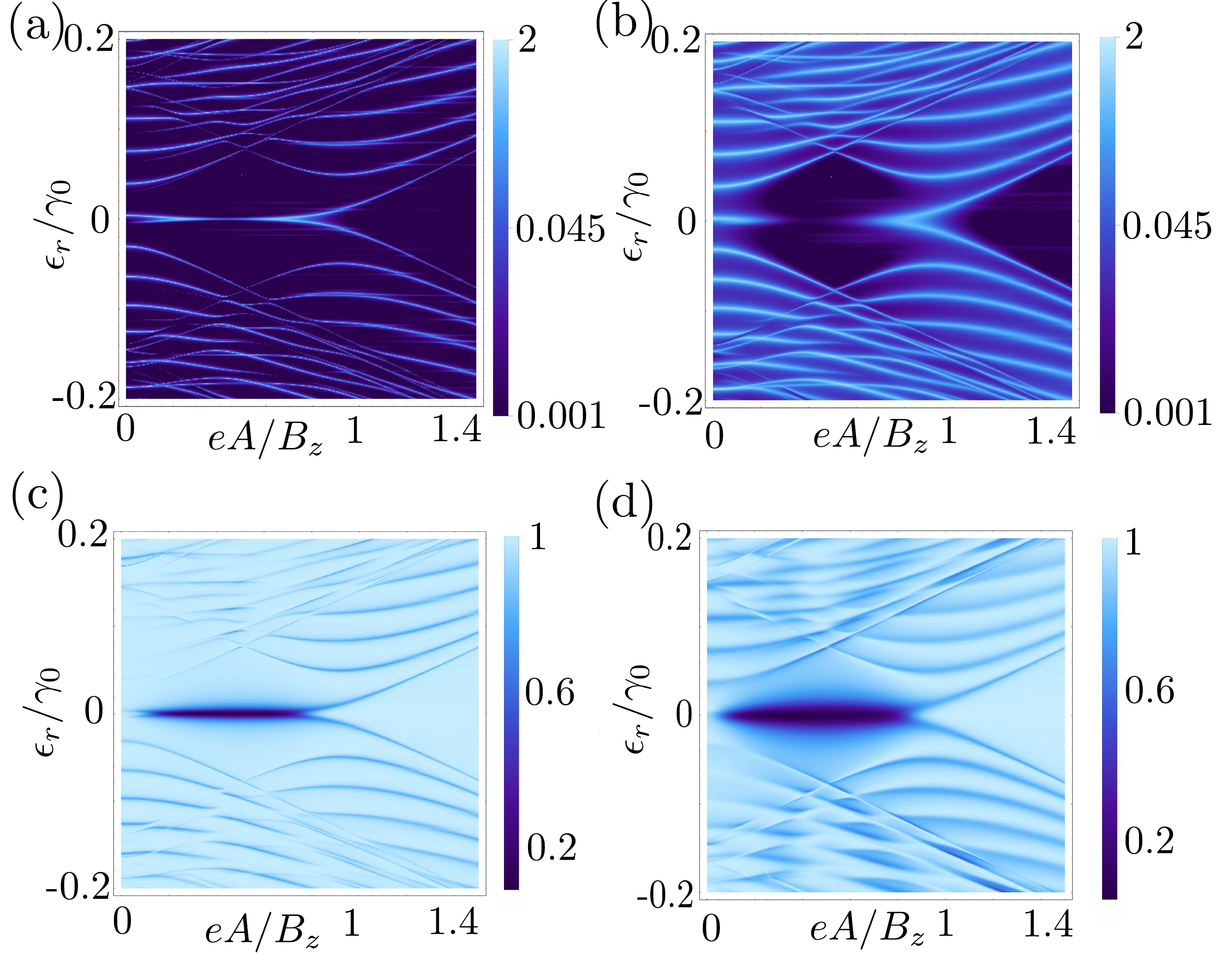}
	\label{Fig:anti_wire}
	\caption{(a-b) Two terminal and (c-d) four terminal conductance (in units of $e^2/h$) through a decoupled QPC under the influence of driven electro-magnetic vector field $eA$. $\epsilon_r = \epsilon-\omega/2$, $\omega = 2\sqrt{\gamma_0^2+\gamma_c^2}$, $L = 120 ~ v_F/\gamma_0$, $L_{\mathrm{bar}}=1/2~ v_F/\gamma_0$, $v_F=1$, $\overline{m}=5$, $\gamma_0=\gamma_c$, $\gamma_0=0.5$, $eA = 0.1$. Further parameter choices are (a) $\lambda/\gamma_0=10$, (b) $\lambda/\gamma_0=6$, (c) $\lambda/\gamma_0=6$, and (d) $\lambda/\gamma_0=4$.}
\end{figure}
The absence of a qualitative difference in the conductance of topological and trivial regime in the two-terminal conductance needs an explanation. The reason for this indistinguishability is inbuilt in the particular coupling to the leads: each helical edge is strongly coupled to the QPC, however, only to half of the channels. This selective coupling directly affects the visibility of the topological phase.

To better understand why, we now look at the complementary detection scheme shown in Fig. \ref{Fig:schematic_supp}.
Additionally to the setup discussed before, we add time reversal invariant barriers of length $L_{\mathrm{bar}}$ and strength $\lambda$ at each end of the QPC. These barriers are described by the Hamiltonian
\begin{equation}
H_{\mathrm{barrier}} =\lambda\bigg[ \int_{-L_{\mathrm{bar}}}^0 dx + \int_{L}^{L+L_{\mathrm{bar}}} dx\bigg] \Psi^{\dagger}(x)\tau_x\sigma_0\Psi(x).
\end{equation} 
They have two effects: the coupling to the leads is suppressed by a factor $\exp(-\lambda L_{\mathrm{bar}})$, and the upper and lower helical edges merge at the barrier. Hence, as far as the two-terminal conductance is concerned, the weak coupling to the leads allows us to resolve the eigenstates in the QPC. The results are shown in Fig. \ref{Fig:anti_wire}. For weak coupling (Fig. \ref{Fig:anti_wire} (a)) the localized eigenstates within the QPC region are clearly visible in the two terminal conductance. In fact, at $eA/B_z=1$, the Floquet gap closes and the system enters the topological phase for $eA/B_z<1$, where a topological mid-gap state is formed. In this regime, the physics is dominated by tunneling processes. This leads to an enhanced transmission, whenever the energy of the incoming particles matches an energy level in the weakly coupled QPC. The visibility of the topological bound state is reduced when the coupling strength is increased (i.e. the barrier strength is lowered) (Fig. \ref{Fig:anti_wire} (b)). In this scenario, the helical nature of incoming particles gains relevance. Then, a prominent bound state signature is found in the conductance of an isolated helical edge as compared to the combined signal of two edges. Indeed, stronger coupling increases the visibility of the topological mid-gap state in the four-terminal conductance, where it appears as a finite reflection into the same channel at the resonant frequency (Fig. \ref{Fig:anti_wire} (c-d)). The reason for this reflection reveals itself when the spin properties of the topological mid-gap state are analyzed, which we explain below.

\begin{figure}[h]
	\centering
	\includegraphics[scale=0.41]{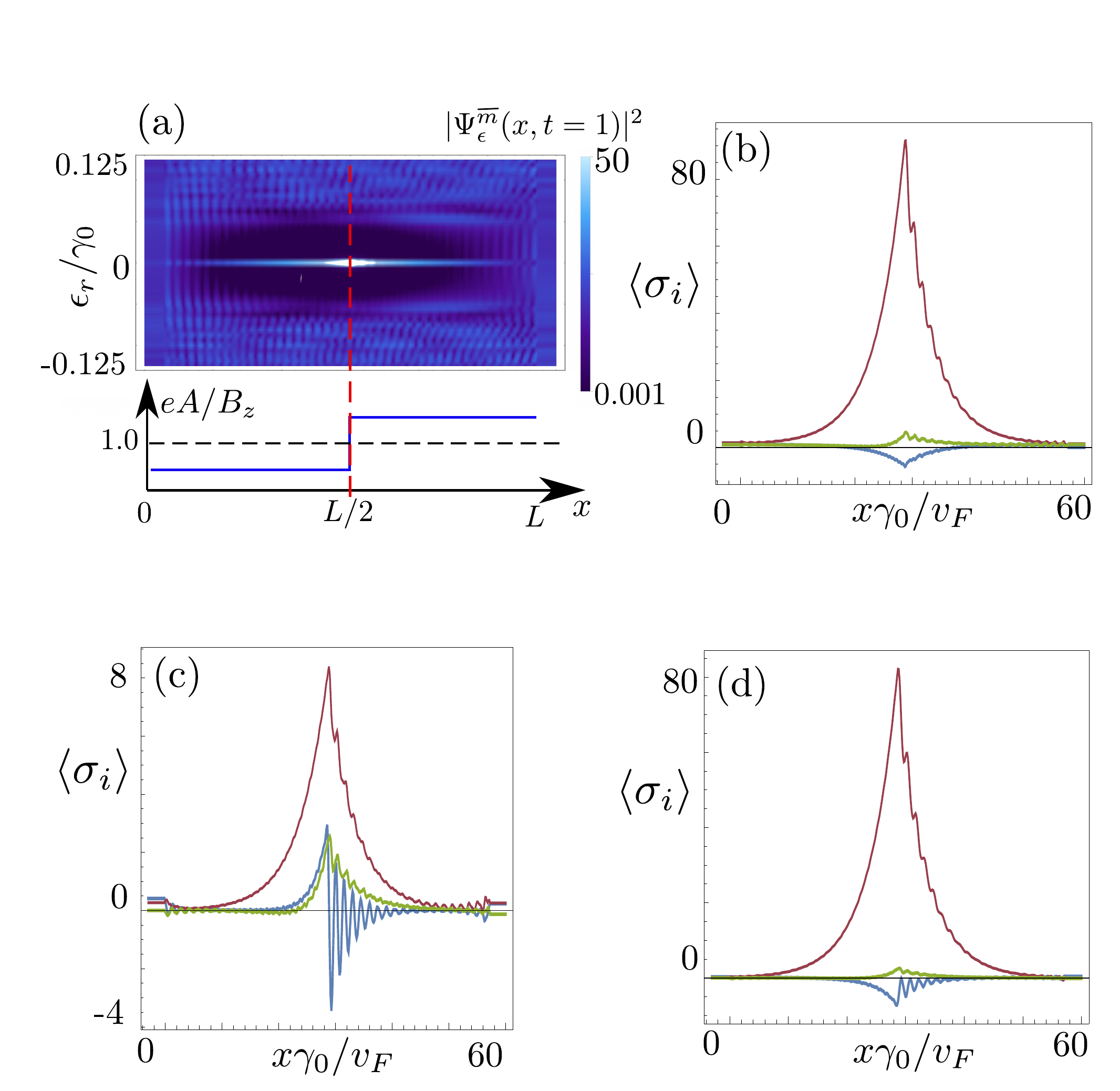}
	\label{Fig:spin}
	\caption{$(a)$ Probability density, in units of $\gamma_0/v_F$ according to Eq. (\ref{Eq:states}) with $\overline{m}=5$, $L=60~ v_F/\gamma_0$, $\epsilon_r=\epsilon-\omega/2$, $\omega=2\sqrt{\gamma_0^2+\gamma_c^2}$ ($\gamma_0=\gamma_c$) and $(eA(x)-B_z(x))/(eA(x))=1/2~\mathrm{sign}(L/2-x)$. $(b-d)$ Expectation value of the spin density, $\langle\sigma_i\rangle$, in units of $\gamma_0/v_F$, with $i\in\{x,y,z\}$ (blue,red,green) evaluated for the system shown in Fig. 3 at quasi energy $\epsilon = \omega/2$ and $\omega = 2\sqrt{\gamma_0^2+\gamma_c^2}$. (b) total spin density of the scattering state. (c-d) spin density  for upper edge (c) and lower edge (d), separately. Other parameters are chosen as in $(a)$.}
\end{figure}

\textit{Bound state.--}The change in the conductance can be directly associated with the presence of a topological bound state at the ends of the QPC. To better visualize the topological state, we consider, within the constricted region, a phase boundary between a topologically gapped and a trivially gapped region in the center of the QPC (at $x=L/2$, see Fig. \ref{Fig:spin}(a)).
The more complex bound state structure corresponding to the setup in Fig.3 is completely analogous and is presented in the SM. More explicitly, we solve the scattering problem related to Eq. (\ref{Eq:eig_prob}) with $(eA(x)-B_z(x))/(eA(x))=(1/2)\mathrm{sign}(L/2-x)$ for the Floquet modes $\phi^{\overline{m}}_{\epsilon,n}(x)$ up to order $\overline{m}$. Then, the Floquet-like solution $\Psi^{\overline{m}}_{\epsilon}(x,t)$ is given by
\begin{equation}
\label{Eq:states}
\Psi^{\overline{m}}_{\epsilon}(x,t)=e^{-i\epsilon t}\sum_{m=-\overline{m}}^{\overline{m}} \phi^{\overline{m}}_{\epsilon,m}(x)e^{-im \omega t}.
\end{equation} 
As we are working within the framework of an open system, we find a solution to the scattering problem for any $\epsilon$. We can name a particular solution a bound state when there is a local maximum that grows as the gap inducing parameters $g$ are increased \cite{invisible}. Then, for the limit $g\rightarrow \infty$, this particular solution becomes a true bound state. Fig. \ref{Fig:spin}(a) shows the probability density of the solutions $\Psi^{\overline{m}}_{\epsilon}(x,t)$ for different $\epsilon$. Note that we have chosen $t=\gamma_0^{-1}$. However, since $\phi_{\epsilon,n}^{\overline{m}}(x)\rightarrow 0$ already for moderately large $n$, the solution depends weakly on time. Moreover, Fig. \ref{Fig:spin}$(a)$ shows the presence of a mid-gap bound state associated with the Floquet topological phase transition. \textcolor{black}{As far as the localization length is concerned\cite{compositemajorana}, there are two relevant length-scales: the Floquet gap dictates the decay of the $2k_F$ components of the bound state, while the components with wavevector close to zero decay proportionally to the inverse of the "topological" gap $|eA-B_z|$. The gap that separates the state from the continuum states is the smallest of the two.}

In Figs. \ref{Fig:spin} (b-d), we depict the expectation value of the different spin density components with respect to the bound state. We recognize a dominant spin component perpendicular to the spin-quantization axis of the helical edge states. Hence, by tunneling into this state, it is possible to flip the spin and thus obtain a finite reflection amplitude. In that sense, the topological bound state plays the role of a ferromagnetic barrier. The prominent process when scattering into this state is hence not resonant tunneling but backscattering into the same channel. This is possible, since time-reversal symmetry is explicitly broken in our system by the applied Zeeman field.

The imbalance of spin density components between the upper and lower edge found in Fig. \ref{Fig:spin} (c-d) can be explained by the different Fermi wavevector of incident particles for $\epsilon=\omega/2$ at the upper, lower edge, respectively. Particles, incident towards the QPC from different edges, mainly see either the induced spectral gap around $k=0$ or $k=k_F$. Since the nature of the gap around $k=k_F$ is different from the one at $k=0$, this eventually manifests in different spin-components of the corresponding part of the bound state. This spin structure explains the backscattering into the same channel for the topological regime discussed above.

\textit{Conclusions.--} We have shown that a Floquet driven QPC between helical edge states represents a feasible solid state setup hosting a Floquet topological quantum phase transition. Moreover, we have demonstrated that photon-assisted transport measurements are able to detect the Floquet topological quantum phase transition and to fully characterize the associated boundary states.
\begin{acknowledgements}
This work was supported by the DFG (SPP1666, SFB1170 “ToCoTronics”),
the W\"urzburg-Dresden Cluster of Excellence ct.qmat,
EXC2147, project-id 39085490, the Elitenetzwerk
Bayern Graduate School on “Topological Insulators” and the Studienstiftung des Deutschen Volkes.
\end{acknowledgements}

\section{Supplementary Material}
In this supplementary material, we provide the explicit derivation of the effective Hamiltonian and the form of the scattering states. We finally discuss the properties of the bound states in the weakly couple quantum point contact.
\subsection{Effective Hamiltonian}
Time periodic Hamiltonians with $H(t)=H(t+T)$ imply solutions of the form $\vert\psi(t)\rangle=\vert\nu(t)\rangle e^{-it\epsilon}$, with $|\nu(t)\rangle=|\nu(t+T)\rangle$. Due to the time-periodicity, $\epsilon$ is only defined $\mathrm{modulo}\,(2\pi/T)$. 
Using $\vert\psi(t)\rangle$ as an ansatz in the time-dependent Schr\"odinger equation $H(t)\vert\psi(t)\rangle=i\partial_t\vert\psi(t)\rangle$, we obtain the eigenvalue equation for the time-dependent states $\vert\nu(t)\rangle$
\begin{eqnarray}
\label{Eq:Q}
Q\vert \nu(t)\rangle=\epsilon\vert\nu(t)\rangle
\end{eqnarray}
with the quasi-energy operator $Q=H(t)-i\partial_t$. Eq. (\ref{Eq:Q}) describes an eigenvalue problem in an extended Hilbert space $\mathcal{F}=\mathcal{H}\otimes\mathcal{L}_T$ constructed by the Hilbert space $\mathcal{H}$ and the space of square-integrable $T$-periodic functions $\mathcal{L}_T$. 
The scalar product defined on $\mathcal{F}$ then combines time averaging with the scalar product on $\mathcal{H}$: $\langle\langle \nu(t)\vert \mu(t)\rangle\rangle = 1/T \int_0^T\mathrm{d}t \langle \nu(t)\vert \mu(t)\rangle$.
A complete set of orthonormal states on $\mathcal{F}$ is given by $\vert\nu_m(t)\rangle=\vert\nu\rangle e^{im\omega t}$ with $\omega=2\pi/T$. Expanding the quasi-energy operator $Q$ in the basis states $\vert\nu_m(t)\rangle$ yields
\begin{eqnarray}
\label{Eq:Qexpansion}
\langle\langle \nu_m(t)\vert Q \vert \mu_n(t)\rangle\rangle =\langle \nu \vert H_{m-n}\vert \mu \rangle+ \delta_{m,n}\delta_{\nu,\mu}m\omega
\end{eqnarray}
with $H_{m}=1/T\int_0^T\mathrm{d}t e^{-im\omega t}H(t).$ Using $H(t)= H_p+H_{\gamma_0}+H_{\gamma_c}+H_B+H_A(t)=H_{QPC}+H_B+H_A(t)$, we obtain only three non-zero contributions
\begin{eqnarray}
\label{Eq:H0}
H_0&=&H_{QPC}+H_B,\\
\label{Eq:H1}
H_1&=&H_{-1}=\frac{1}{2}\int_0^L\mathrm{d}x\Psi^{\dagger}(x)\left(-eA\tau_z\sigma_z\right)\Psi(x).
\end{eqnarray}
Explicitly, the Fermi field operators in the constricted region read $\Psi(x)=\sum_k \textbf{c}_k e^{ikx}$ with creation operators $\textbf{c}_k=(\hat{c}_{k1R},\hat{c}_{k1L},\hat{c}_{k2L},\hat{c}_{k2R})^T$. We now apply a unitary transformation $\textbf{U}$ such that $H_{\mathrm{QPC}}$ becomes diagonal. The new creation operators $\tilde{ \boldsymbol d}_k=\textbf{U}\textbf{c}_k$ are then associated with the eigenvalues $\pm E(\pm k)=\pm \sqrt{\gamma_0^2+(\gamma_c\pm kv_F)^2}$ (Fig. 1 (b) of the main text). In particular, we associate the fermions $\hat{d}_k^{\pm \pm}$ with the eigenvalue $\pm E(\pm k)$. In terms of the transformed fermions $\tilde{\boldsymbol d}_k$, $H_B$ and $H_{\pm 1}$ become
\begin{eqnarray}
H_B\!&=&\!\sum_k\!\tilde{\boldsymbol d}^{\dagger}_k\!\big[\!B_{0}(k)\tau_0\sigma_x\!+\!\frac{B_1(k)(\tau_x\sigma_x\!\!-\!\tau_y\sigma_y)}{2}\nonumber\\
&+&\!\frac{B_2(k)(\tau_x\sigma_x\!+\!\tau_y\sigma_y)}{2}\big] \tilde{\boldsymbol d}_k,~\\
H_1\!&=&\!\sum_k \tilde{\boldsymbol d}^{\dagger}_k\big[\frac{\Delta_0(k)\tau_x(\sigma_0-\sigma_z)}{2}+\frac{\Delta_0(-k)\tau_x(\sigma_0+\sigma_z)}{2}\nonumber\\
&+&\frac{\Delta_1(k)(\tau_z\sigma_0-\tau_z\sigma_z)}{2}\!-\!\frac{\Delta_1(-k)(\tau_z\sigma_0\!+\!\tau_z\sigma_z)}{2}\big]\!\tilde{\boldsymbol d}_k~~~~~~
\end{eqnarray}
with $\Delta_0(k)=-eA\vert \gamma_0\vert/\sqrt{\gamma_0^2+(\gamma_c+kv_F)^2}$ and $\Delta_1(k)=-eA(\gamma_c+kv_F)/\sqrt{\gamma_0^2+(\gamma_c+kv_F)^2}$. The explicit $k$-dependence of $B_0~,B_1$ and $B_2$ is too lengthy to be presented here. These parameters obey $B_{0},~B_1,~B_2\propto B_z$. In the resonant case with $\omega=2\sqrt{\gamma_c^2+\gamma_0^2}$ (Fig. 1 (c)), degenerate points in the eigenvalue spectrum of the quasi-energy operator (the ones that are coupled by the arrow and the corresponding ones for $k>0$ in Fig. 1 (c) of the main text) appear. Thus, matrix elements related to the coupling between these eigenvalues become dominant. Moreover, provided $eA,B_z\ll\omega$, matrix elements describing the coupling between states that are separated by at least $\omega$ can be safely neglected for quasi-energies close to the resonance point so to obtain the effective operator given in the main text, where $\textbf{d}_k=(\hat{d}_k^{+-},\hat{d}_k^{++},\hat{d}_k^{--},\hat{d}_k^{-+})$.

\subsection{Floquet scattering states}
In this section, we investigate the scattering states, used to solve the scattering problem given in the main text.

Valid incoming states from the left with $x<0$, as well as a valid outgoing states for $x>L$ are given by
\begin{eqnarray}
\label{Eq:ansatz1}
\psi_{\mathrm{in,{\overline{m}}}}^\alpha(x)\!&=&\!M_{\overline{m}}^p(x,0)\!\bigg[D_{\overline{m},\alpha}\!+\!\mathbb{1}_{\!\overline{m}\!\times\! \overline{m}\!}\,\frac{(\tau_0\sigma_0-\tau_z\sigma_z)}{2} S_{\overline{m}}\bigg],~~~~\\
\label{Eq:ansatz2}
\psi_{\mathrm{out,\overline{m}}}(x)&=&M_m^p(x,0)\bigg[\mathbb{1}_{\overline{m}\times \overline{m}}\frac{(\tau_0\sigma_0+\tau_z\sigma_z)}{2} S_{\overline{m}}\bigg]
\end{eqnarray} 
with
\begin{eqnarray}
\label{Eq:S}
S_{\overline{m}}&=&(t_{1}^{\overline{m}},r_{1}^{\overline{m}},r_{2}^{\overline{m}},t_{2}^{\overline{m}},
\dots,t_{1}^{-\overline{m}},r_{1}^{-\overline{m}},r_{2}^{-\overline{m}},t_{2}^{-\overline{m}})^T,~~\\
D_{\overline{m},\alpha}&=&\sum_{l=1}^4 \delta_{l,\alpha^2}\hat{e}^{(2\overline{m}+1)}_{\overline{m}+1}\otimes \hat{e}^{(4)}_l,
\end{eqnarray}
where $\hat{e}^{(4)}_l$ are Cartesian basis vectors associated to the operators $\textbf{c}_k$, while $\hat{e}^{(2\overline{m}+1)}_{\overline{m}+1}$ is associated to photon space. $\alpha\in \{1,2\}$ is the lead index. The matrix $M_{\overline{m}}^p(x,0)$ is the transfer matrix calculated from the system without QPC potentials , i.e. 
\begin{eqnarray}
M_{\overline{m}}^p(x,0)=\exp\left[\int_0^x\mathrm{d}x\frac{i}{v_F}
P_{\overline{m}}^{-1}(\epsilon-Q_{\overline{m}}^p)\right],
\end{eqnarray}
with the matrix elements $[Q_{\overline{m}}^p]_{k,l}=\delta_{k,l} k \omega $ for $\vert k \vert \leq \vert {\overline{m}}\vert$.
\subsection{Bound state in the strong impurity setup}
\begin{figure}
	\centering
	\includegraphics[scale=0.27]{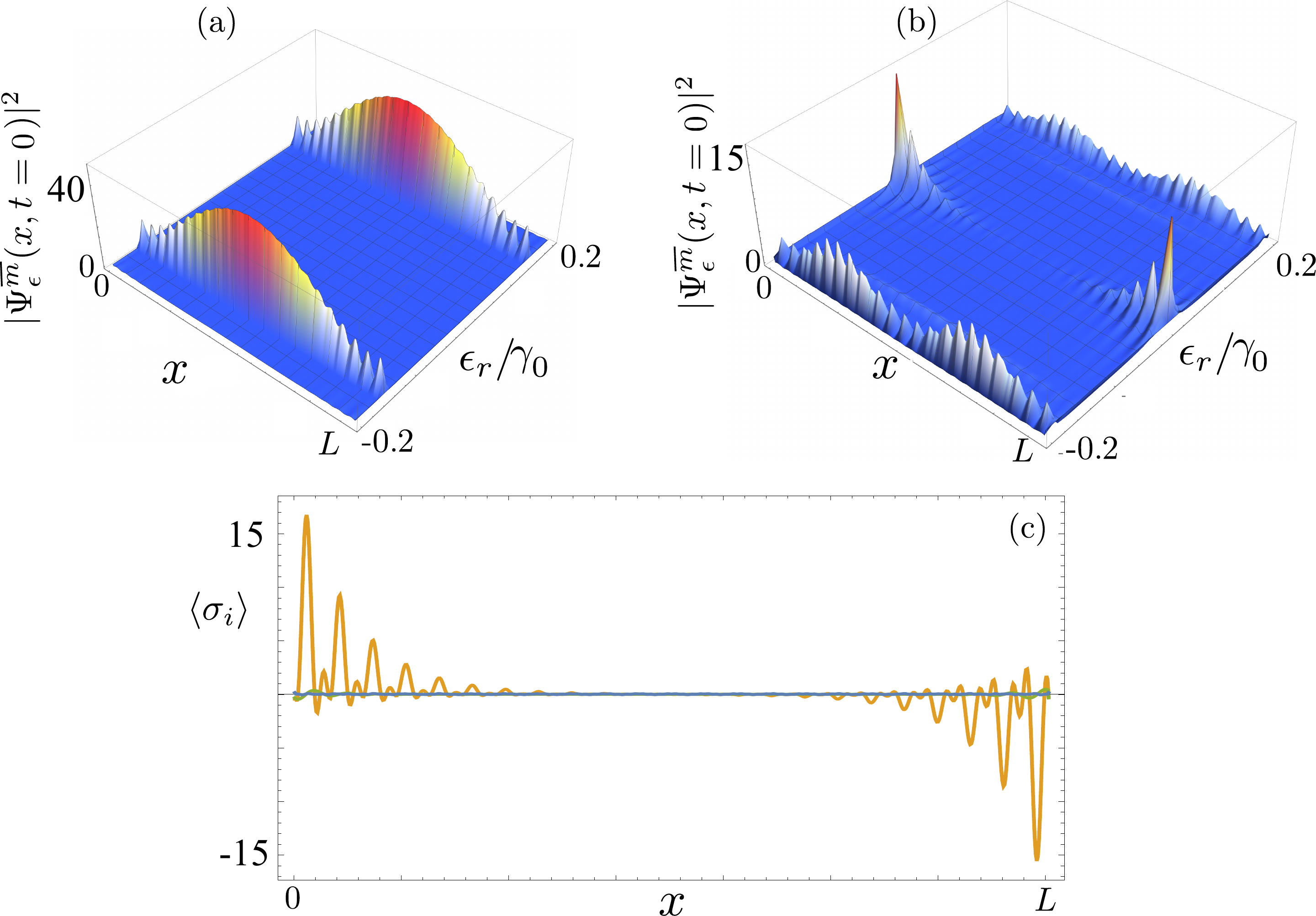}
	\caption{(a) Probability density according to Eq.13 of the main text, in units $\gamma_0/v_F$, as a function of position and quasienergy $\epsilon_r=\epsilon-\omega/2$ with $\omega =2 \sqrt{\gamma_0^2+\gamma_c^2}$ for  $\overline{m} = 5$, $\lambda = 5\gamma_0$, $eA = 0.2\gamma_0$, $B = 0.1\gamma_0$, $\gamma_c=\gamma_0$ and $L=35 v_F/\gamma_0$(b) Same as (a), with $\overline{m} = 5$, $\lambda = 5\gamma_0$, $eA = 0.2\gamma_0$, $B = 0.2\gamma_0$, $\gamma_c=\gamma_0$ and $L=35 v_F/\gamma_0$ (c) Spin densities, green, orange and blue for the $x,y,z$ components respectively, in units $\gamma_0/v_F$, as a function of position. Other parameters as in panel (b).}
	\label{Fig:sup}
\end{figure}
In the setup of Fig.3 of the main text, no bound states are present in the trivial phase (Fig. \ref{Fig:sup} (a)), while two bound states, one at each end of the QPC, emerge in the topological phase (Fig. \ref{Fig:sup} (b)). The spin texture is analogous to the one discussed in the main text for the simplified setup with one bound state only, with $\langle\sigma_y\rangle=\Psi^{\overline{m}*}_{\epsilon_r}(x,0)\tau_0\sigma_y\Psi^{\overline{m}}_{\epsilon_r}(x,0)$ dominating over the other projections.


\begin{thebibliography}{50}
\bibitem{note}Generalizing the above discussion to a driving frequency $\omega=2(\sqrt{\gamma_0^2+\gamma_c^2}-\delta\omega)$, with an imbalance $\delta\omega$, from Eq. (\ref{Eq:Heff}), we obtain a gap closing-reopening transition at $k=0$ for 
\begin{eqnarray}
\label{Eq:GapClosing}
B_z=\pm \sqrt{(eA)^2+(\delta\omega)^2(1+(\gamma_c/\gamma_0)^2)}.
\end{eqnarray}
\bibitem{note2}This assumption is valid, provided the Fermi wave length $\lambda_F$ is much smaller than the smoothening length $L_s$ that describes the build-up of the QPC potentials which itself should be smaller than $L$. Hence, we require $\lambda_F\ll L_s\ll L$.
\bibitem{note3}The sum is due to the fact that an incoming electron can be reflected with the absorption of, generically, $m$ photons.
\end{thebibliography}
\end{document}